\documentclass[pdflatex,sn-mathphys-num]{sn-jnl}

\usepackage{graphicx}%
\usepackage{multirow}%
\usepackage{amsmath,amssymb,amsfonts}%
\usepackage{amsthm}%
\usepackage{mathrsfs}%
\usepackage[title]{appendix}%
\usepackage{xcolor}%
\usepackage{textcomp}%
\usepackage{manyfoot}%
\usepackage{booktabs}%
\usepackage{algorithm}%
\usepackage{algorithmicx}%
\usepackage{algpseudocode}%
\usepackage{listings}%
\usepackage{hyperref}%

\theoremstyle{thmstyleone}%
\theoremstyle{thmstyletwo}%

\theoremstyle{thmstylethree}%

\begin{document}

\title[Article Title]{Anomalously long-delayed afterpulses in large-area photomultipliers}

\author*[1]{\fnm{Nikita} \sur{Ushakov}}\email{nikitaushakoff@gmail.com}

\author[1]{\fnm{Bayarto} \sur{Lubsandorzhiev}}

\author[1]{\fnm{Arslan} \sur{Lukanov}}

\author[1]{\fnm{Andrei} \sur{Sidorenkov}}

\author[1]{\fnm{Dmitrii} \sur{Voronin}}

\affil*[1]{\orgname{Institute for Nuclear Research of the Russian Academy of Sciences}, \orgaddress{\street{pr. 60-letiya Oktyabrya 7a}, \city{Moscow}, \postcode{117312}, \country{Russian Federation}}}


\abstract{We report the observation of anomalously long-delayed afterpulses in photomultipliers of the Baksan Large Neutrino Telescope project~--- 10-inch R7081-100, 8-inch R5912-100, 20-inch R12860 photomultipliers produced by Hamamatsu Photonics, and 20-inch N6205 photomultipliers produced by NNVT. The mean delay times relative to the main pulses are approximately 85~$\mu$s, 73~$\mu$s, 260~$\mu$s, and 90~$\mu$s, respectively. The probability of such afterpulses does not exceed 0.1\% per photoelectron, and their amplitudes are strictly confined to the single-photoelectron level, regardless of the amplitude of the main pulse. The delay time of these afterpulses shows no significant dependence on the PMT operating voltage.}


\maketitle

\section{Introduction}\label{sec1}
Afterpulses are a well-known phenomenon inherent in all classical vacuum photomultipliers (PMTs). They were first observed in the late 1940s and early 1950s~\cite{Godfrey,Mueller}. Since then, numerous studies have been devoted to investigating afterpulses and their origin; we refer the reader to reviews~\cite{Coates_1973_1,Coates_1973_2,Staubert,Yamashita,Torre} for further details. Currently, there is a generally accepted model describing the origin of afterpulses. According to this model, afterpulses originate mainly from ionic feedback, i.e., ionization of residual gas atoms by photoelectrons and secondary electrons in electron multiplication systems such as dynodes or microchannel plates (MCPs)~\cite{Coates_1973_1,Coates_1973_2}. The delay times of afterpulses relative to the main pulses span a wide range from about a hundred nanoseconds to 10--20 microseconds, depending on the ion masses and the PMT size. Afterpulses with delays greater than 20~$\mu$s cannot be readily explained within the ion-feedback model. Glukhovskoy and Yaroshenko~\cite{Glukhovskoy} proposed an exoelectronic emission model to describe such afterpulses. However, there are still many uncertainties in building a comprehensive model of afterpulses to explain all details of the phenomenon.

In 1997, one of us (B.L.) observed unusual afterpulses in 8-inch EMI9350-series PMTs produced by ETL. These afterpulses exhibited a broad, pronounced peak in their delay time distribution between 70 and 120~$\mu$s after the main pulse. However, the results were published in a dedicated paper only 15 years later, in 2012~\cite{Poleshchuk}. The rate of such afterpulses was less than 0.1\% per primary photoelectron, and their amplitudes were strictly confined to the single-photoelectron level. We refer to these events as long-delayed afterpulses. These afterpulses appear anomalous and cannot be readily explained within the ion-feedback model. To date, no other reports have been published on such long-delayed afterpulses, except for information (L. Koepke, private communication) about a possible observation of such events in the IceCube experiment with subsequent publication~\cite{IceCube}.

Understanding the origin and characteristics of long-delayed afterpulses is crucial for large-scale neutrino detectors. These detectors rely, in part, on the inverse beta decay (IBD) reaction, which produces a prompt positron signal followed by a delayed neutron-capture event. The neutron-capture time typically ranges from tens to hundreds of microseconds, depending on the capture medium and the presence of neutron absorbers. This interval overlaps with the delay range of the observed long-delayed afterpulses (40--250~$\mu$s). Consequently, these afterpulses could mimic genuine neutron signals or produce accidental coincidences with preceding positron events, thereby degrading the sensitivity to neutrino oscillations, supernova neutrinos, or other rare processes. Their characterization is therefore essential for reliable background suppression and event reconstruction.

A campaign to study large-area photomultipliers for the new Baksan Large Neutrino Telescope project was recently conducted~\cite{Ushakov:2020,Ushakov:2021}. At the current stage of the project, two types of PMTs produced by Hamamatsu Photonics are used for the prototype detector with a target mass of 5~t: 10-inch R7081-100 as the main PMTs and 8-inch R5912-100 for the water Cherenkov detector. The 20-inch Hamamatsu R12860 PMTs and the NNVT N6205 MCP-PMTs have also been studied as candidate PMTs for a full-scale detector. During the characterization of these PMTs, long-delayed afterpulses were observed in all tested devices. This paper presents the results of this study.

\section{The setup}\label{sec2}
\begin{figure}[h]
\centering
\includegraphics[width=0.7\textwidth]{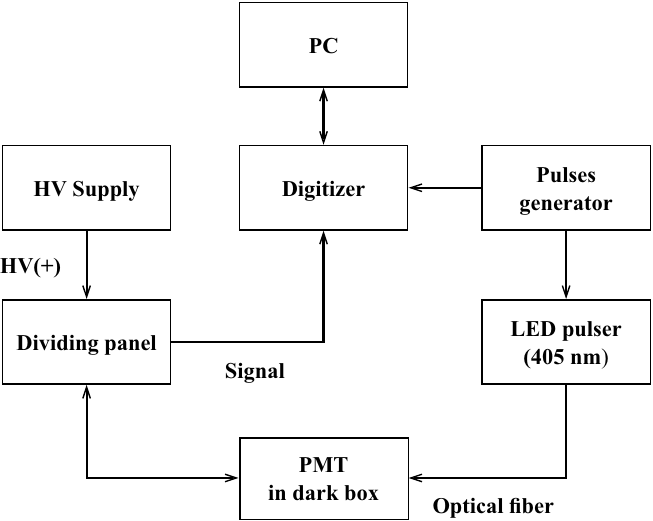}
\caption{The setup used for PMT characteristics measurements.}
\label{fig1}
\end{figure}

To measure the afterpulses, a standard experimental setup was used, including a digitizer and an LED synchronized with it by a pulse generator. Figure~\ref{fig1} shows a block diagram of the measurement setup. As a digitizer, we used the CAEN V1730 with a sampling rate of 500~MS/s. A 405~nm LED was used as the light source. Custom-developed software was used to measure the amplitudes and delays of all pulses following the main pulse. The threshold for pulse detection for each PMT was set to approximately 0.25~PE (photoelectrons). The acquisition window was chosen such that the long-delayed afterpulse peak extended down to the background level determined by dark current pulses. To avoid the influence of the Earth's magnetic field on the PMT, a magnetic field compensation system was used~\cite{Lukanov}.

\section{Afterpulse studies}\label{sec3}

\begin{figure*}[ht]
\begin{minipage}[b]{0.5\textwidth}
\centering
\includegraphics[width=1\textwidth]{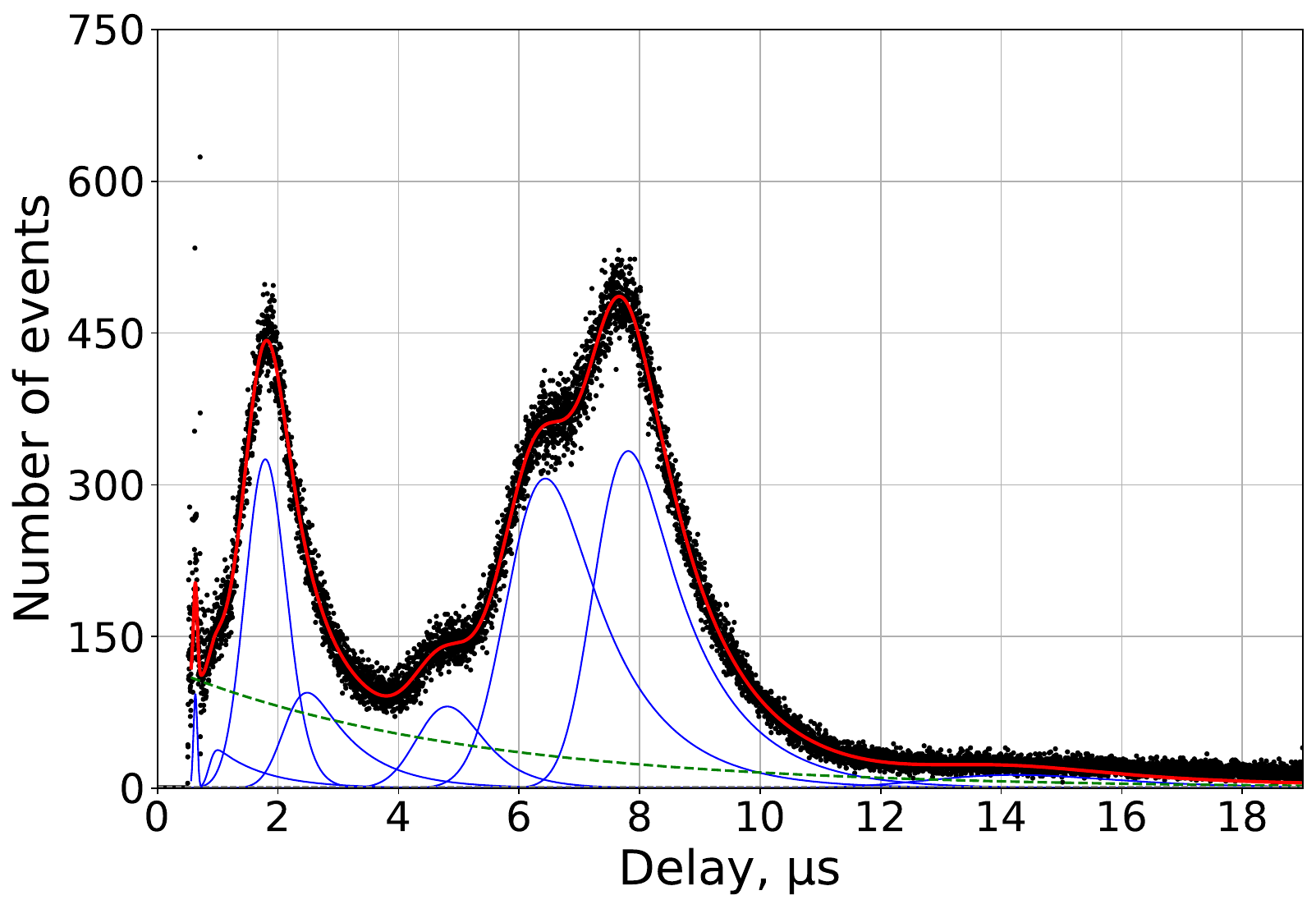}
\end{minipage}
\begin{minipage}[b]{0.5\textwidth}
\centering
\includegraphics[width=1\textwidth]{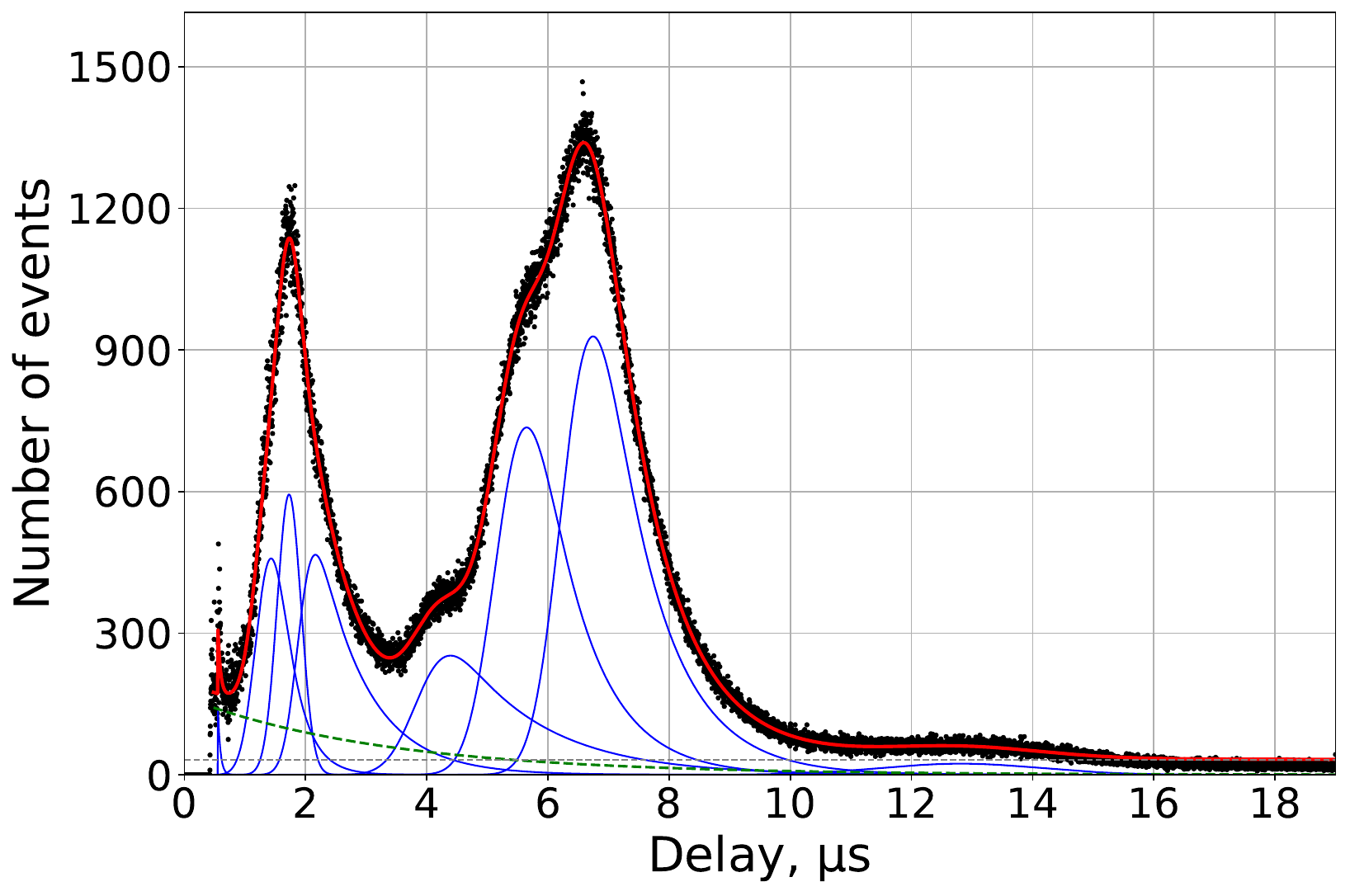}
\end{minipage}
\caption{Typical delay distribution of afterpulses for the PMTs R7081-100 (left) and R5912-100 (right). The red curves show the total best fit, and the blue curves show individual fit components.}
\label{fig2}
\end{figure*}

We first examine classical afterpulses, taking R7081-100 and R5912-100 PMTs as examples. A batch of fifty R7081-100 and twelve R5912-100 photomultipliers was characterized at a gain of about $10^7$ under illumination corresponding to several tens of PE. Figure~\ref{fig2} shows a typical delay distribution of afterpulses for the entire batch. The red curves show the total best fit, and the blue curves show the individual fit components, exponentially modified Gaussians (EMGs).

The possible ion species were identified by comparing the observed afterpulse delays with the expected ion transit times~\cite{Coates_1973_1,Incandela}. Following the commonly used approximation for large hemispherical photomultipliers, the potential distribution between the photocathode and the first dynode was assumed to have a quadratic dependence. Assuming that ions are created near the first dynode with negligible initial kinetic energy, integration of the equation of motion gives the ion transit time
\begin{equation}
\Delta t=\frac{\pi}{4}\sqrt{\frac{2 m}{q V_0}}L,
\end{equation}
where $V_0$ is the potential difference between the photocathode and the first dynode and $L$ is the corresponding distance, $q$ and $m$ are the ion charge and mass, respectively. $L$ is about 16~cm for R7081-100 and about 13~cm for R5912-100. For the case shown in Figure~\ref{fig2}, the potential differences $V_0$ were 635~V and 658~V, respectively.

Based on these results, the possible origin of each peak was identified. For example, for the R7081-100 PMT, afterpulses can be attributed to the following ions: 0.63~$\mu$s ($\rm H^+$), 1~$\mu$s ($\rm H_2^+$), 1.8~$\mu$s ($\rm He^+$), 2.5~$\mu$s ($\rm C^+,~O^+,~CH_4^+$), 4.81~$\mu$s ($\rm CO_2^+$), 6.45~$\mu$s ($\rm Kr^+$), 7.81~$\mu$s ($\rm Xe^+$), and 14.26~$\mu$s ($\rm CsO_2^+$). For individual PMTs, the values may vary slightly depending on the gas concentration and the operating voltage.

\begin{figure*}[ht]
\begin{minipage}[ht]{0.5\textwidth}
\centering
\includegraphics[width=1\textwidth]{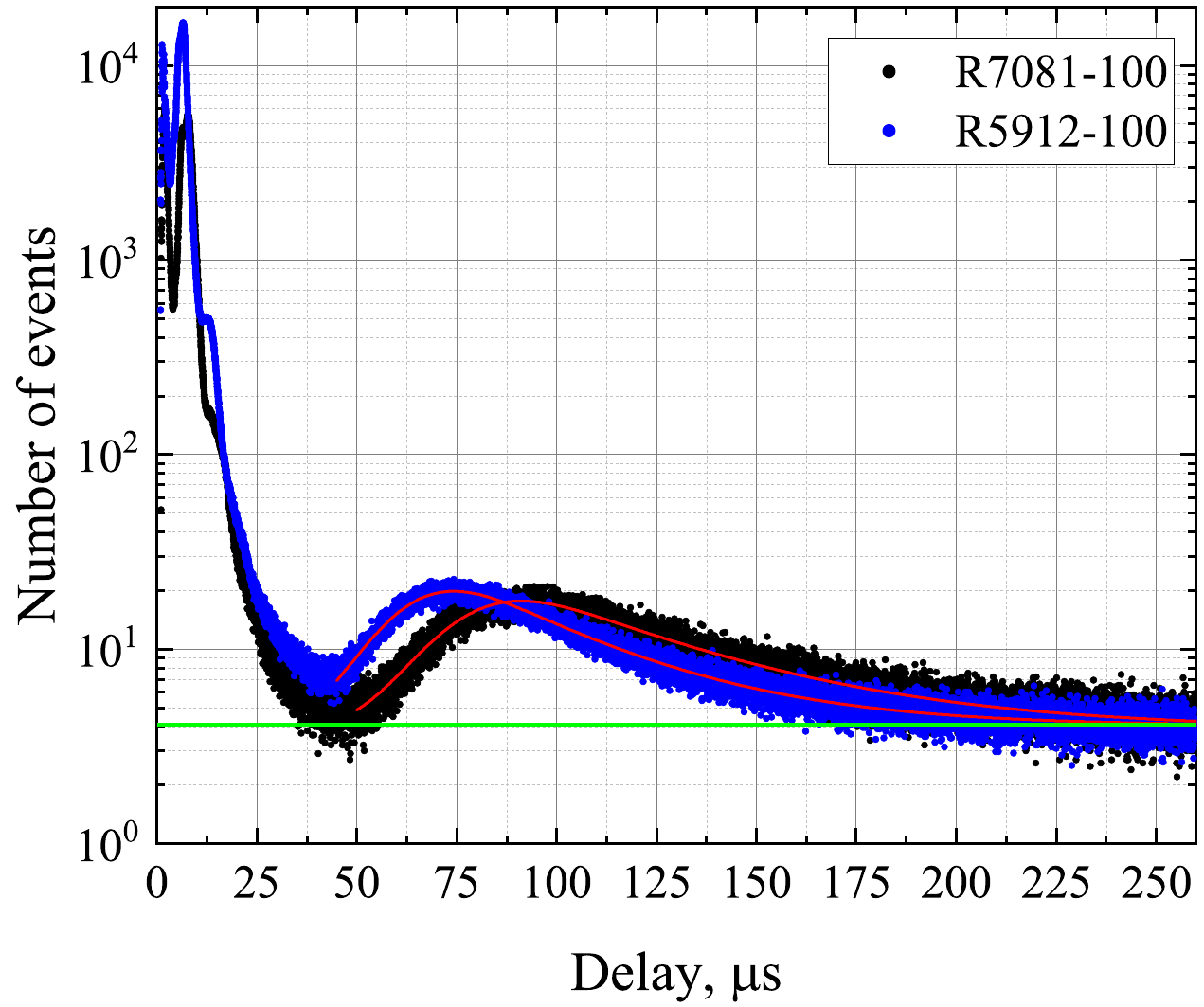}
\end{minipage}
\begin{minipage}[ht]{0.5\textwidth}
\centering
\includegraphics[width=1\textwidth]{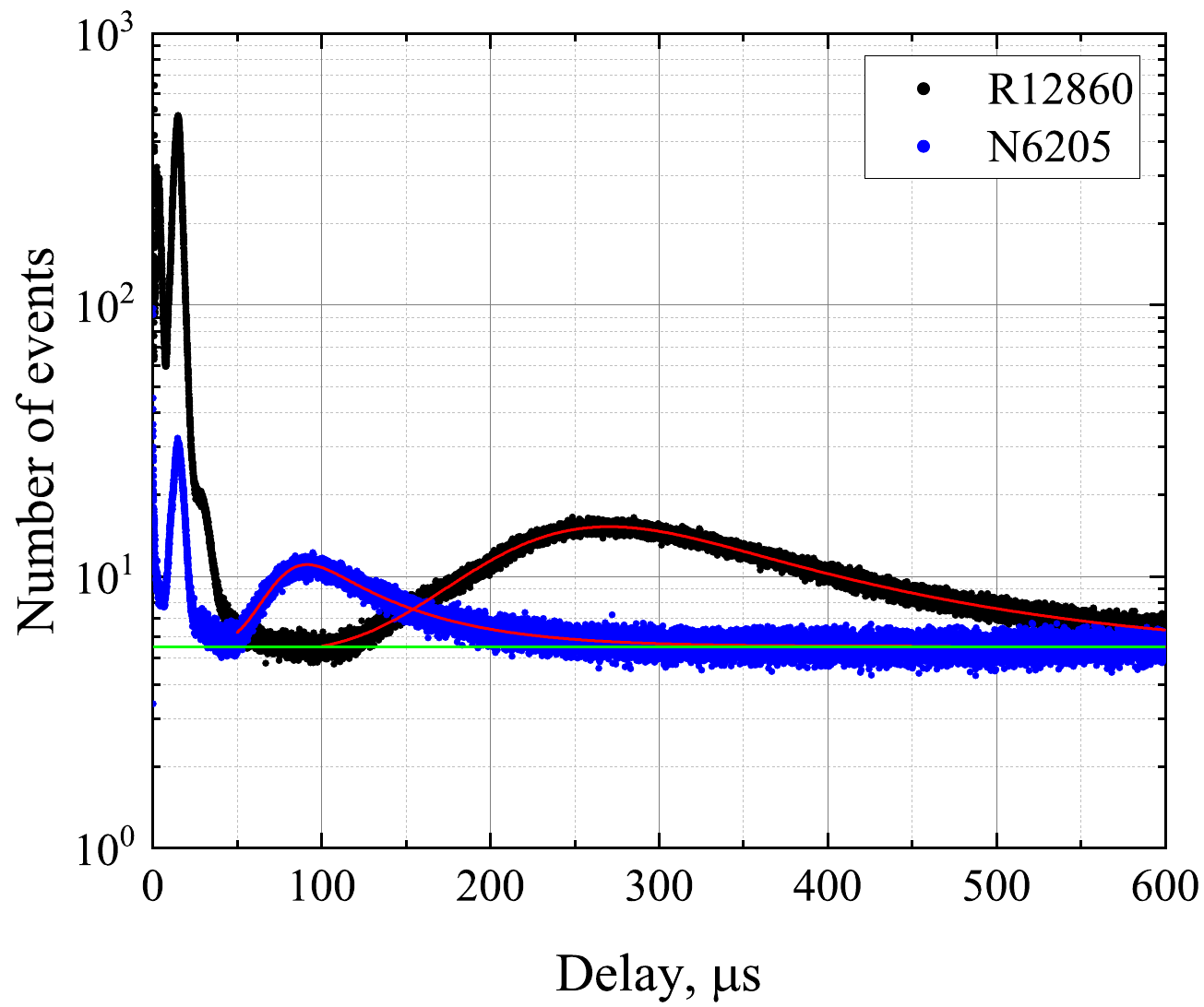}
\end{minipage}
\caption{Typical delay distributions of long-delayed afterpulses. Left: R7081-100 (black) and R5912-100 (blue). Right: R12860 (black) and N6205 (blue). In both panels, the red curves show the fitting function and the green curves show the noise level.}
\label{fig3}
\end{figure*}

After increasing the illumination to several hundred PE, in addition to the classical afterpulses, a broad, pronounced peak of the long-delayed afterpulses was observed for all the R7081-100 PMTs in the range 50--250~$\mu$s and the R5912-100 PMTs in the range 45--230~$\mu$s. The positions of these peaks are approximately $85\pm 5~\mu$s and $73\pm 3~\mu$s, respectively. The left panel of Figure~\ref{fig3} shows typical delay distributions of long-delayed afterpulses for the PMTs R7081-100 and R5912-100, with the red line showing the fitting function (EMG), and the green line showing the background level due to dark current noise. Similar measurements were carried out for one R12860 PMT and one N6205 PMT. The right panel of Figure~\ref{fig3} shows delay distributions of long-delayed afterpulses for these PMTs; for the R12860 PMT, the delay time of these afterpulses is in the range 100--600~$\mu$s with the peak in the delay distribution at approximately 260~$\mu$s, while for the N6205 PMT the corresponding range is 50--250~$\mu$s and the peak is at approximately 90~$\mu$s.

\begin{figure*}[ht]
\begin{minipage}[ht]{0.5\textwidth}
\centering
\includegraphics[width=1\textwidth]{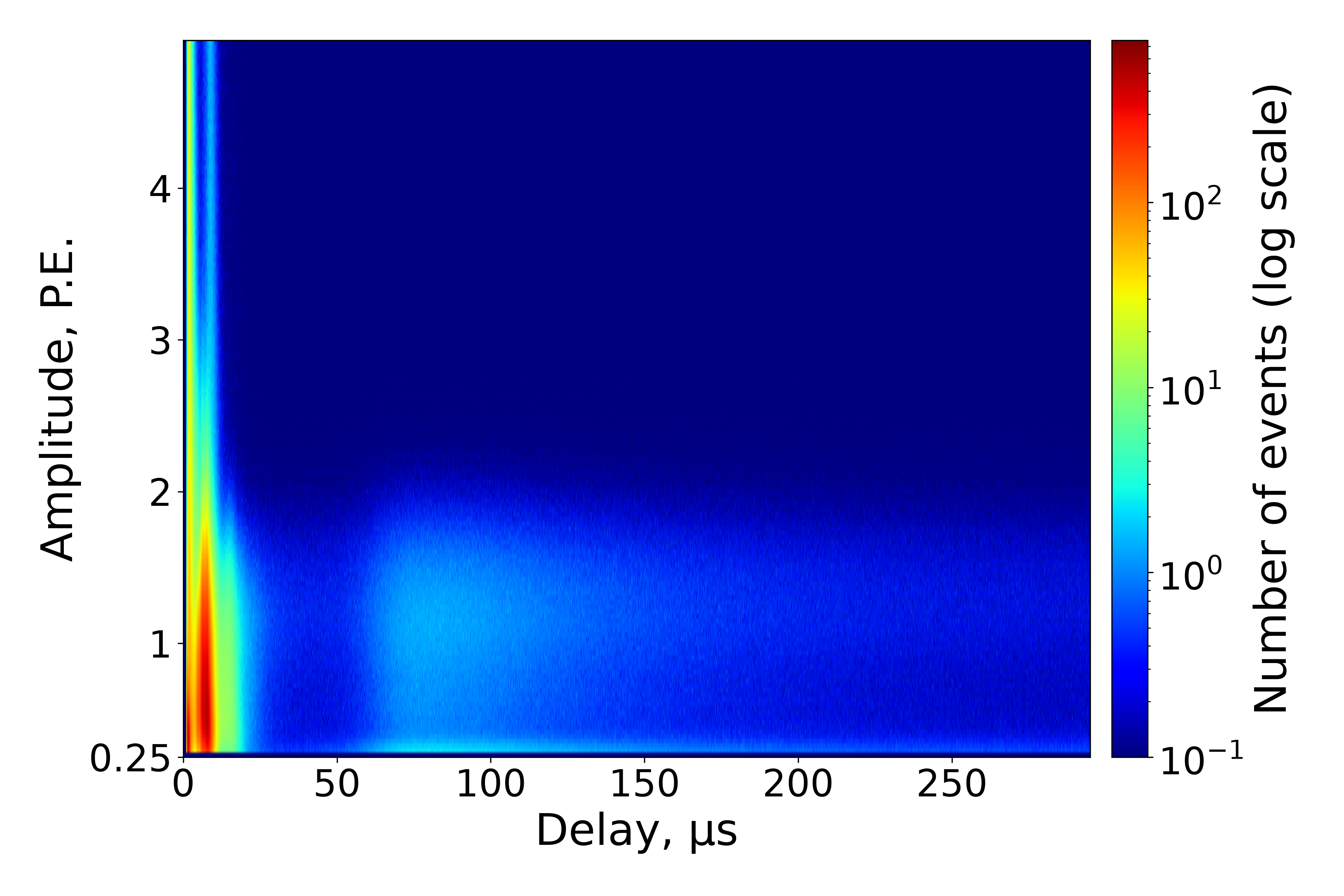}
\end{minipage}
\begin{minipage}[ht]{0.5\textwidth}
\centering
\includegraphics[width=1\textwidth]{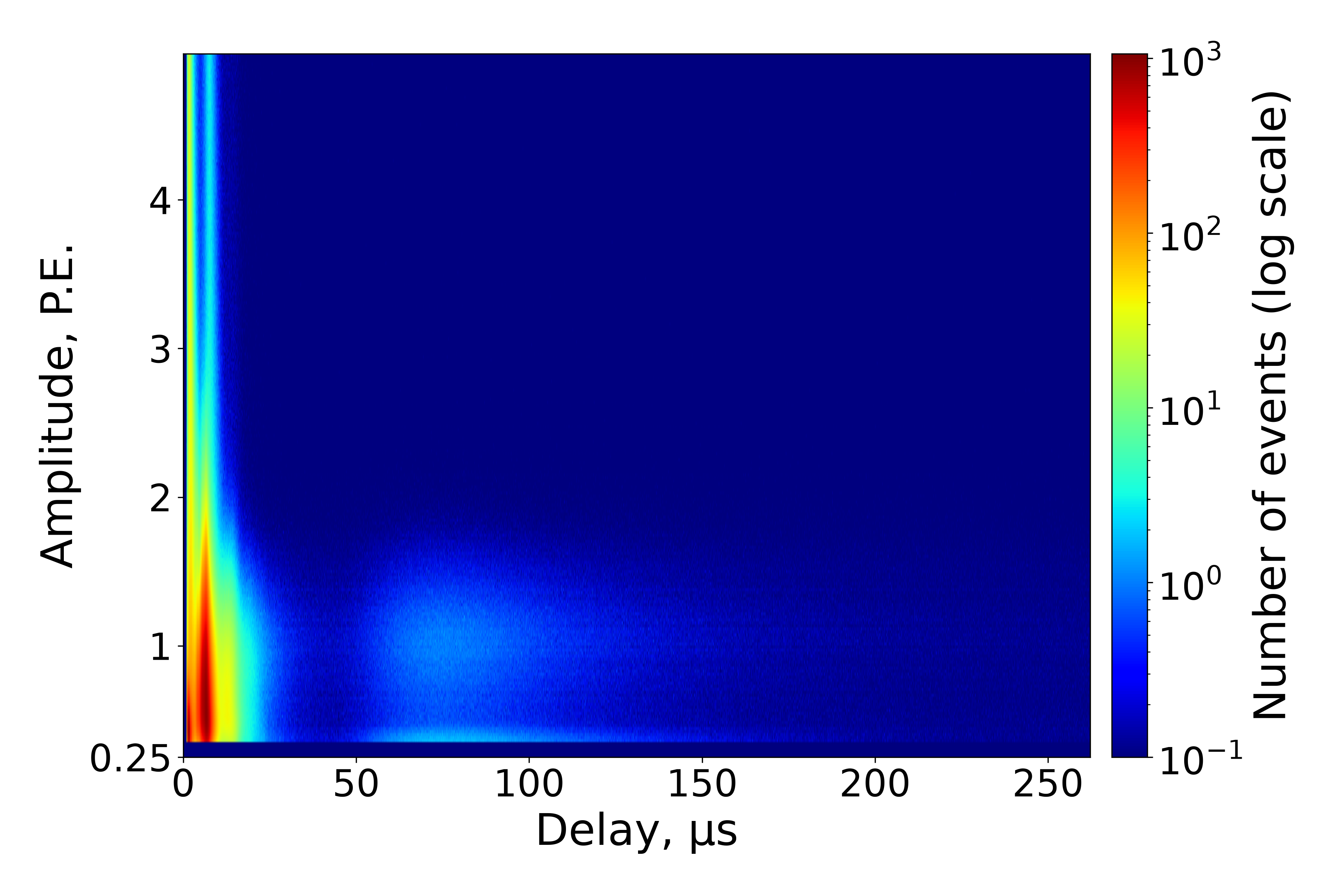}
\end{minipage}
\caption{Typical amplitude-delay distributions of long-delayed afterpulses for the PMTs R7081-100 (left) and R5912-100 (right).}
\label{fig4}
\end{figure*}

\begin{figure*}[ht]
\begin{minipage}[ht]{0.5\textwidth}
\centering
\includegraphics[width=1\textwidth]{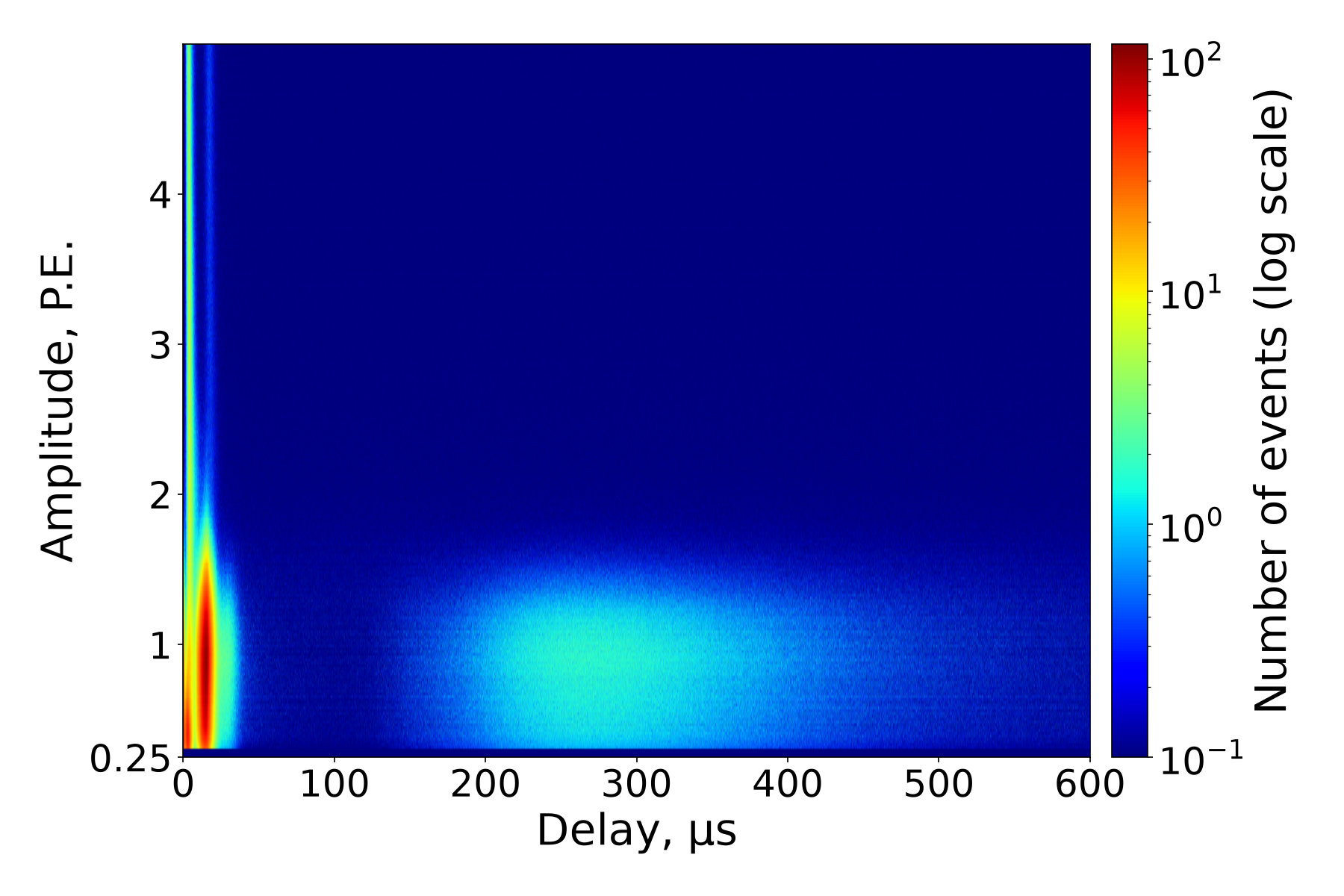}
\end{minipage}
\begin{minipage}[ht]{0.5\textwidth}
\centering
\includegraphics[width=1\textwidth]{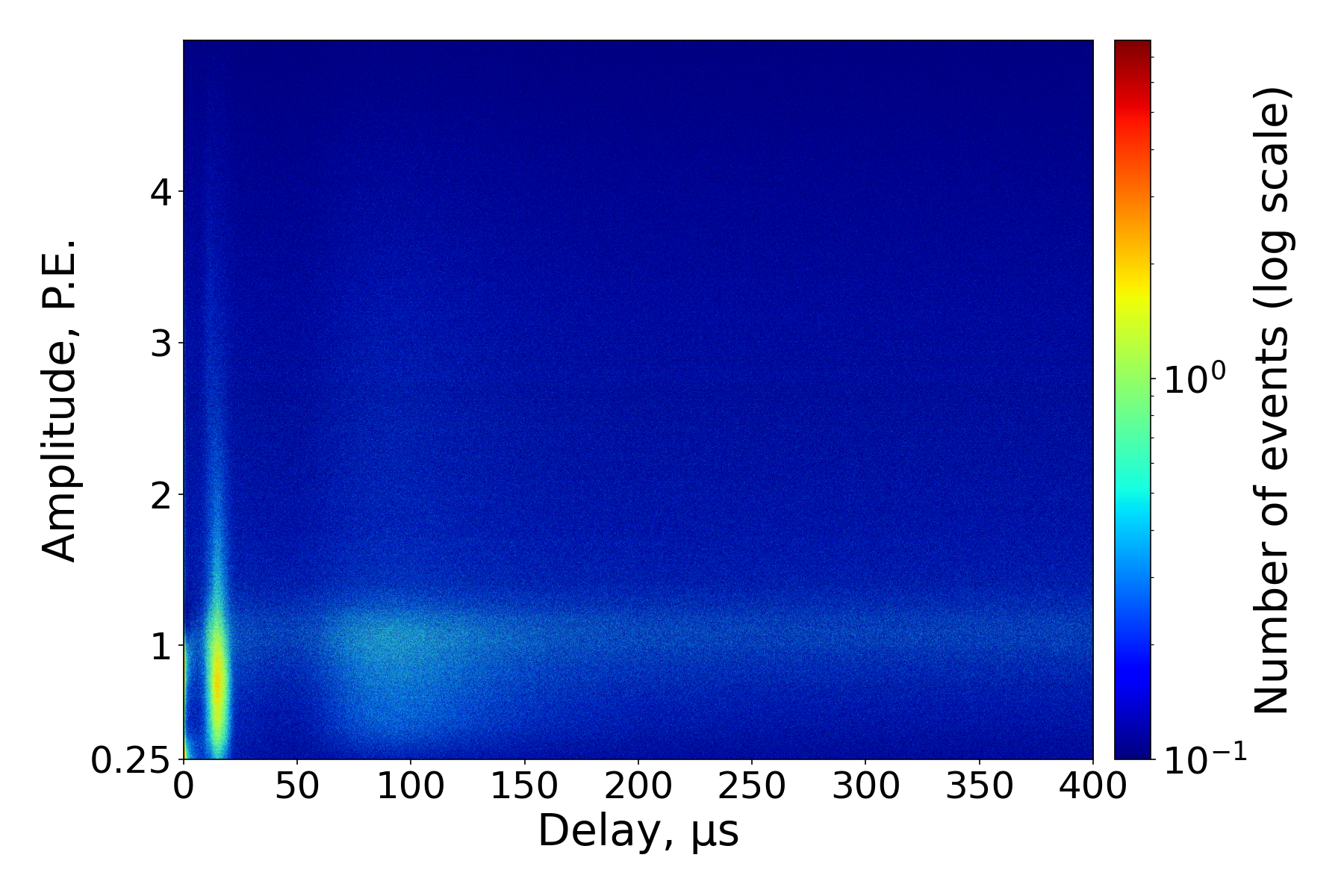}
\end{minipage}
\caption{The amplitude-delay distributions of long-delayed afterpulses for the PMTs R12860 (left) and N6205 (right).}
\label{fig5}
\end{figure*}

The source of this peak cannot be identified using the method described above, since the molecular mass of the corresponding ion would be greater than 10,000~u, which is physically unrealistic for any ion originating from residual gases. The probability of their occurrence is in the range 0.005--0.1\% per photoelectron, depending on the sample. Figure~\ref{fig4} shows a typical amplitude-delay distribution of long-delayed afterpulses for the R7081-100 and R5912-100 PMTs; Figure~\ref{fig5} shows similar distributions for the R12860 and the N6205 PMTs. The measured distributions show that these afterpulses have a single-photoelectron amplitude.

\section{Operating voltage dependence}\label{sec4}
\begin{figure}
\centering
\includegraphics[width=0.7\textwidth]{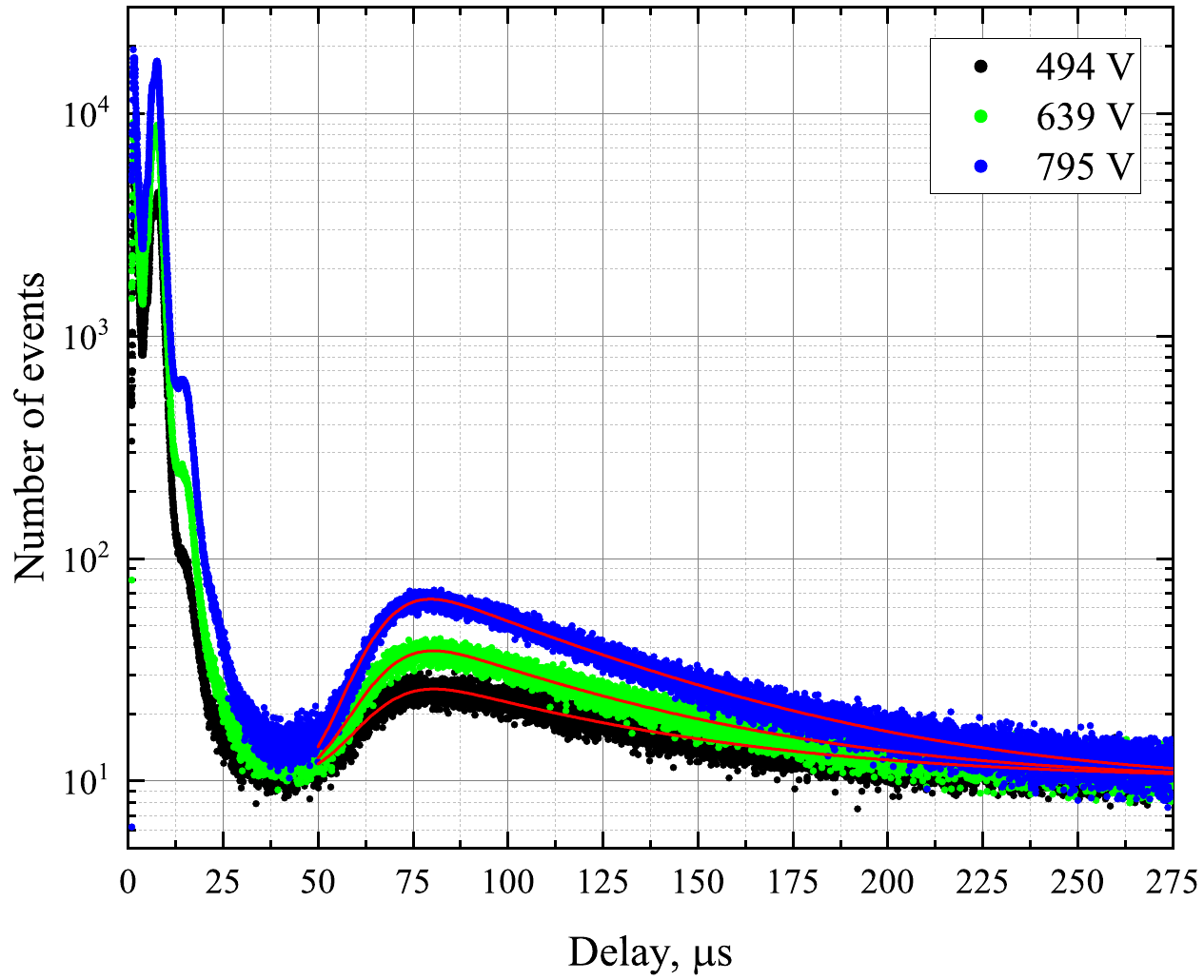}
\caption{Delay distributions of long-delayed afterpulses for the R7081-100 PMT at potential differences between the photocathode and the first dynode of 494~V (black), 639~V (green), and 795~V (blue) and the corresponding fits (red curves).}
\label{fig6}
\end{figure}

To investigate the possible influence of the applied high voltage on the delay time of long-delayed afterpulses, we performed dedicated measurements with a representative R7081-100 PMT. Measurements were performed at three different operating voltages of 1230, 1590, and 1980~V, corresponding to gains of $1.3\times10^6$, $10^7$, and $5\times10^7$, and to potential differences between the photocathode and the first dynode of 494, 639, and 795~V, respectively. For each setting, the delay distributions of the long-delayed afterpulses were recorded under identical illumination conditions (several hundred photoelectrons). Figure~\ref{fig6} shows the delay distributions of long-delayed afterpulses and their fits at different voltages.

The mean delay times showed no significant variation with the voltage. Specifically, the centroid of the afterpulse peak remained at approximately 80~$\mu$s for the R7081‑100 across the tested voltage range. This absence of voltage dependence further distinguishes these afterpulses from conventional ion‑feedback afterpulses, whose transit times typically scale with the inverse square root of the accelerating voltage. However, the probability of occurrence of long-delayed afterpulses, like that of classical afterpulses, varies among individual PMTs.

\section{Conclusion}\label{sec5}
The anomalously long-delayed afterpulses were observed in the 10-inch R7081-100, 8-inch R5912-100, 20-inch R12860 PMTs produced by Hamamatsu Photonics, and the 20-inch N6205 PMT produced by NNVT. The delay times relative to the main pulse are in the range 50--250~$\mu$s ($mean\approx85\pm 5~\mu$s) for R7081-100, 45--230~$\mu$s ($mean\approx73\pm 3~\mu$s) for R5912-100, 100--600~$\mu$s ($mean\approx260~\mu$s) for R12860, and 50--250~$\mu$s ($mean\approx90~\mu$s) for N6205. The results indicate a clear correlation between the afterpulse delay time and the photomultiplier size. No clear dependence of the afterpulse delay time on the operating voltage of the PMTs was observed. Their amplitudes are strictly confined to the single-photoelectron level. The probability of these long-delayed afterpulses is below 0.1\% per photoelectron. The observed afterpulses do not appear to be explainable within the conventional ion-feedback model, and further studies are needed to shed light on their origin.

\bmhead{Acknowledgements}
This work is supported by the Ministry of science and higher education of the Russian Federation under the contract No.\,075-15-2020-778 in the framework of the Large scientific projects program within the national project "Science".

\bibliography{sn-bibliography}

\end{document}